\documentclass[aip,cha,reprint]{revtex4-1}

\usepackage[pdftex]{graphicx}
\usepackage{amssymb,amsfonts,amsmath}
\usepackage{grffile}
\usepackage{hyperref}

\begin{document}

%\preprint{}

\title{Chimeras in globally coupled oscillatory systems: From
  ensembles of oscillators to spatially continuous media} %Title of paper



\author{Lennart Schmidt}
%\thanks{}
\affiliation{Physik-Department, Nonequilibrium Chemical Physics, Technische Universit\"{a}t M\"{u}nchen,
  James-Franck-Str. 1, D-85748 Garching, Germany}
\affiliation{Institute for Advanced Study - Technische Universit\"{a}t M\"{u}nchen,
  Lichtenbergstr. 2a, D-85748 Garching, Germany}

\author{Katharina Krischer}
\email[]{krischer@tum.de}
\affiliation{Physik-Department, Nonequilibrium Chemical Physics, Technische Universit\"{a}t M\"{u}nchen,
  James-Franck-Str. 1, D-85748 Garching, Germany}

\date{\today}

\begin{abstract}
We study an oscillatory medium with a nonlinear global coupling that 
gives rise to a harmonic mean-field oscillation with constant
amplitude and frequency. Two types 
of cluster states are found, each undergoing a symmetry-breaking 
transition towards a related chimera state. We demonstrate that the 
diffusional coupling is non-essential for these complex dynamics. 
Furthermore, we investigate localized turbulence and discuss whether
it can be categorized as a chimera state.
\end{abstract}


\maketitle %\maketitle must follow title, authors, abstract and \pacs

\begin{quotation}
An oscillatory medium may exhibit a variety of spatio-temporal patterns. 
A cluster pattern
consisting of distinct uniformly oscillating regions with mutual phase shifts is
a prominent example. 
Surprisingly, the symmetry of this state can be broken
and a so-called chimera state may form. Then, some of the regions
display turbulent dynamics, while the other regions remain
synchronized. This coexistence of synchrony and incoherence arouse
much interest and triggered many investigations. Unihemispherical
sleep, performed by, e.g., various dolphins and birds, may be connected
to the phenomenon of chimera states, since during this kind of sleep
one cerebral hemisphere is awake (desynchronized), while the other one
is sleeping (synchronized).

The prerequisites for such chimera states are not completely known
yet. For globally coupled systems it seems that a clustering mechanism
is needed as a first symmetry-breaking step. In this Article, we
investigate two different kinds of cluster states that undergo each a
transition to a corresponding chimera state. The chimera states then
share properties with the cluster states, from which they
originate. Furthermore, we discuss the role of the diffusional
coupling and point out that in many cases it is sufficient to consider
only the global coupling in order to understand the complex dynamics. However, so-called
localized turbulence seems to rely on diffusion. This state is similar
to chimera states, while the lack of a phase boundary between
synchrony and incoherence renders its detailed dynamics different.
\end{quotation}

%\section{}
%\label{}
%\subsection{}
%\subsubsection{}

\section{Introduction}

Oscillatory media with a global coupling display a wealth of
spatio-temporal dynamics. Considering the medium as a field of
diffusively coupled oscillators \cite{Kuramoto_2003}, each oscillator experiences exactly
the same force from the global coupling. Counterintuitively, this can
lead to a bistability of the dynamics, i.e., there exist two
stable stationary states and each oscillator can settle to one of
them. Therefore, cluster formation
may be observed. In a cluster state, the system separates into
distinct regions with mutual phase shifts. In several experimental
setups a global coupling arises naturally or can be implemented easily and thus, cluster formation
has been observed in various systems \cite{Vanag_Nature_2000, 
Vanag_JPCA_2000, Mikhailov_PhysicsReports_2006, Varela_PCCP_2005, 
Miethe_PRL_2009}.

Astonishingly, the symmetry of
the cluster state can be broken in such a way that one observes the
coexistence of synchronously oscillating regions and regions
displaying turbulence. This state has been termed a chimera state
\cite{AbramsStrogatz_PRL_2004}, referring to the chimera in Greek 
mythology, a hybrid creature consisting of parts of different animals. Chimera states 
have been found in many different settings. Most of the theoretical 
investigations deal with phase models with a nonlocal coupling scheme, 
i.e., the coupling strength decreases with the distance between two 
oscillators \cite{Kuramoto_NPCS_2002, AbramsStrogatz_PRL_2004, 
Panaggio_Nonlinearity_2015}. Furthermore, discrete systems like 
coupled map lattices, may exhibit chimera states. Finally, chimera states have 
been demonstrated to exist in several experiments. For a recent 
review, see Ref.~\cite{Panaggio_Nonlinearity_2015}.

The photoelectrodissolution of n-type silicon is one of the 
experimental systems exhibiting chimera states. Unlike in most 
settings, in this system the chimera states emerge spontaneously, without
external feedback or specifically prepared initial conditions
\cite{Schmidt_Chaos_2014, Schoenleber_NJP_2014}. On the silicon working electrode an oxide 
layer forms and the thickness of this layer features the 
spatio-temporal patterns. We modeled this continuous medium 
successfully employing a complex Ginzburg-Landau equation with a 
nonlinear global coupling \cite{Schmidt_Chaos_2014, 
Schmidt_chapter_ECCII}. Omitting the diffusional coupling in the equation 
and thus, studying an ensemble of Stuart-Landau oscillators with 
nonlinear global coupling, we were able to demonstrate that the 
chimera states we found form under solely global coupling 
\cite{Schmidt_Chaos_2014}. A more detailed investigation of chimera 
states in the Stuart-Landau ensemble - in fact, we found two types of 
chimeras - revealed that for globally coupled systems a clustering 
mechanism and non-isochronicity of oscillators is sufficient for 
chimera states to exist \cite{Schmidt_PRL_2015}. Non-isochronicity means that the frequency of 
oscillation depends on the actual amplitude. These conclusions are in 
accordance with other studies on chimera states under global coupling 
\cite{Yeldesbay_PRL_2014, Chandrasekar_PRE_2014}.

In this Article, we bring continuous media and ensembles of oscillators 
closer together by demonstrating that both types of chimera states 
found in the Stuart-Landau ensemble with nonlinear global coupling are 
present in the modified complex Ginzburg-Landau equation (MCGLE), the 
model for the photoelectrodissolution of n-type silicon. More 
precisely, we present two types of cluster states giving rise to two 
types of chimera states and connect them with the corresponding states 
in the discrete ensemble of oscillators. It turns out that the 
mechanisms discussed for the discrete system are also valid in the 
continuous model. Especially, the chimera state found in the ensemble 
under linear global coupling \cite{Sethia_PRL_2014} is also present in 
the corresponding CGLE with linear global coupling. Furthermore, we 
investigate also so-called localized turbulence, for which diffusion 
seems to be essential, and discuss whether this state should also be 
classified as a chimera state.

\section{CGLE with nonlinear global coupling}
In the photoelectrodissolution of n-type silicon, an oxide-layer 
forms at the Si$|$electrolyte interface. Its thickness is determined by the interplay 
of electrooxidation and an etching process and can exhibit sustained oscillations 
\cite{Miethe_PRL_2009, Miethe_PhDTh}. In fact, these oscillations 
originate in a Hopf bifurcation, as shown in 
Refs.~\cite{Miethe_JEC_2012, Schmidt_chapter_ECCII}. When spatially resolving the
oxide-layer thickness, one observes that it can also exhibit spatio-temporal 
pattern formation with a huge variety of different dynamical states 
\cite{Miethe_PRL_2009, Schoenleber_NJP_2014}. Astonishingly, for most of
the patterns, the spatially averaged oxide-layer thickness oscillates 
almost harmonically. To model these dynamics, i.e., a 
system near a supercritical Hopf bifurcation with conserved harmonic mean-field 
oscillations, in a general way, we use a complex Ginzburg-Landau equation with an 
additional nonlinear global coupling \cite{Miethe_PRL_2009, 
GarciaMorales_PRE_2010, Schmidt_Chaos_2014, Schmidt_chapter_ECCII}:
\begin{align}
  \partial_t W = &W + (1+i c_1)\nabla^2 W  - (1+i c_2)|W|^2 W \notag \\
  & - (1+i \nu)\left< W \right> + (1+i c_2)\left< |W|^2 W 
  \right> \ .
  \label{eq:MCGLE}
\end{align}
$W$ describes a complex two-dimensional field $W=W(x,y,t)$, where $x$ 
and $y$ are spatial coordinates and $t$ denotes time. $\left< \dots 
\right>$ stands for spatial averages. Calculating the spatial average 
of the whole equation, we obtain for the mean-field
\begin{equation}
 \partial_t \left< W \right> = -i \nu \left< W \right> \ \Rightarrow \
 \left< W \right> = \eta e^{-i \nu t} \ .
 \label{eq:MCGLE_avg}
\end{equation}
Thus, we indeed describe a two-dimensional system in the vicinity of a 
Hopf bifurcation with conserved harmonic mean-field oscillation. We 
solve Eq.~\eqref{eq:MCGLE} numerically as described in the Appendix. 
In the next two sections we examine two types of cluster patterns
undergoing each a symmetry-breaking transition towards a chimera state.

\section{Type I dynamics}
\subsection{Nonlinear global coupling}

\begin{figure*}[h!]
  \centering
  \includegraphics[width=145mm]{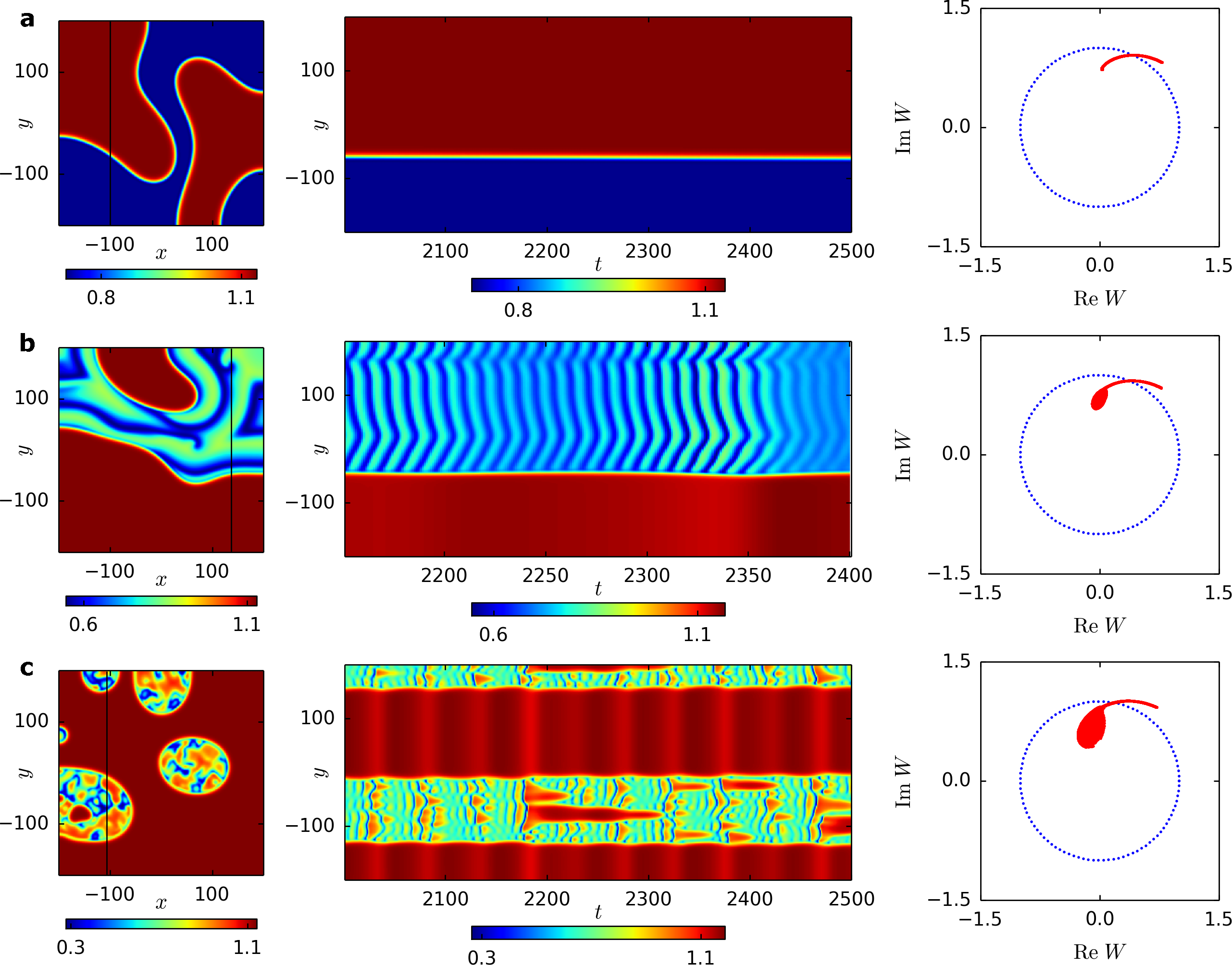}
  \caption{Type I patterns found in Eq.~\eqref{eq:MCGLE}: snapshots of $\left| W \right|$ (left column),
    one-dimensional cuts versus time (also $\left| W \right|$) as indicated by the vertical lines in the snapshots (middle column) and snapshots of the
    arrangement of local oscillators in the complex plane (right column). (a) Amplitude clusters
    ($c_1=0.2$, $c_2 = 0.56$, $\nu = 1.5$, $\eta = 0.9$). 
    (b) Coexistence of synchrony and spiral-wave like dynamics. Waves are
    emitted from the region boundaries and from the spiral cores ($c_1=0.2$, $c_2 =
    0.58$, $\nu = 1.5$, $\eta = 0.9$). (c) Type I chimera state
    consisting of oval-shaped regions exhibiting desynchronized
    behavior in an otherwise homogeneous background
    ($c_1=0.2$, $c_2 = 0.61$, $\nu = 1.5$, $\eta = 1.0$).}
\label{fig:typeI}
\end{figure*}

The first type of dynamics we present are patterns
related to well-known amplitude clusters \cite{Daido_PRL_2006}. The dynamics of such
amplitude clusters are visualized in Fig.~\ref{fig:typeI}a in a
two-dimensional snapshot showing $\left| W(x,y) \right|$ (left), a one
dimensional cut as indicated in the snapshot versus time (middle) and
a snapshot of the arrangement of the local oscillators in the complex
plane (right). Parameters read $c_1=0.2$, $c_2 = 0.56$, $\nu = 1.5$
and $\eta = 0.9$. Amplitude clusters consist of regions separated
mainly by an amplitude difference. Both groups oscillate, with a small 
phase difference to each other, at constant, but different amplitude.

Changing parameter $c_2$ to $c_2 = 0.58$, one of the cluster regions
undergoes a symmetry-breaking transition, as shown in Fig.~\ref{fig:typeI}b. Synchronized regions of
constant amplitude coexist with regions, where amplitude waves are
emitted from the boundaries and from amplitude-spiral cores. The
amplitude spirals are visible in the two-dimensional
snapshot. Interestingly, there is no amplitude defect in the spiral
core and the spiral dynamics take place in a very curved and confined
region.
Here, dynamics are mainly found in the modulus, while the phase
hardly exhibits any spatial pattern and is more or less uniform in the
two regions, with a small phase shift between the regions. This is
also obvious from the arrangement in the complex plane. In the
amplitude cluster state in Fig.~\ref{fig:typeI}a, the two groups of
oscillators reside at the two ends of the string visible in the complex
plane. Going to spiral-wave like dynamics in Fig.~\ref{fig:typeI}b,
the group at the lower radius spreads and thus forms a
bunch. Inside the bunch the relative amplitude differences are much
larger than the mutual phase differences.

This bunch of oscillators at the lower radius desynchronizes strongly
when changing parameters to $c_2 = 0.61$ and $\eta = 1.0$. Now, the
symmetry-breaking is more dramatic: Synchronized regions of constant
amplitude coexist with regions displaying amplitude turbulence,
see Fig.~\ref{fig:typeI}c. This is an example for a chimera state,
which is mainly present in the modulus of the dynamics. Dynamics in the
desynchronized regions are reminiscent of intermittent behavior in the
standard complex Ginzburg-Landau equation
\cite{Chate_Nonlinearity_1994}: homogeneous spots pop up in an
irregular manner, vanishing slowly afterwards. Amplitude defects do
not occur. One observes a slight oscillation in the amplitude of the homogeneous part
and this seems to be connected to variations in the spatial extension of the
desynchronized regions, as visible in the one-dimensional
cut. Furthermore, in both, the spiral-wave like dynamics and the
chimera state, the dynamics in the incoherent regions are overlayed by
an overall oscillation, washing out the patterns repeatedly, as
visible, e.g., at the end of the cut shown in Fig.~\ref{fig:typeI}b. However,
it does not seem that the dynamics are resetted, but rather the
visualization becomes blurry.

Type I chimeras are related to amplitude clusters, as the 
two groups in both states are separated by an amplitude difference. 
Thus, the clustering mechanism is needed to yield two separated groups 
in order to obtain the dynamics presented here. Then, the symmetry is 
broken due to nonlinear amplitude effects: since the response on a 
force depends on the amplitude of the oscillator, the response is 
different in the two groups at different moduli.
As we will see in
Section~\ref{sec:typeII}, the modulated amplitude clusters give
rise to a second type of chimera states.

\subsection{Linear global coupling}
\begin{figure*}[ht!]
  \centering
  \includegraphics[width=\textwidth]{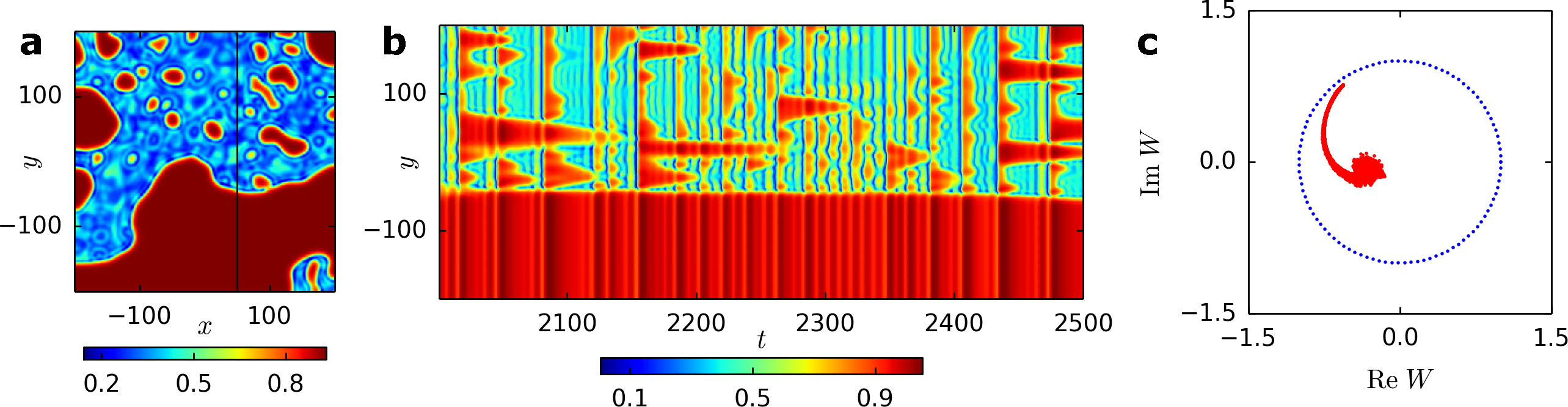}
  \caption{Type I chimera state in a CGLE with linear global
    coupling, given in Eq.~\eqref{eq:CGLE_LGC}. Shown are a 
    two-dimensional snapshot of the modulus (a), a
    one-dimensional cut of the modulus versus time (b) and the arrangement of
    local oscillators in the complex plane (c). Parameters read $c_1 =
    1.2$, $c_2 = 1.7$, $K = 0.67$ and $c_3 = -1.25$.}
\label{fig:LGC_chimera}
\end{figure*}

%\clearpage
%
Interestingly, one finds type I chimeras also in a CGLE with linear
global coupling:
\begin{align}
  \frac{\partial W}{\partial t} &= W - (1+ic_2) \left| W \right|^2 W +
  (1+ic_1) \nabla^2 W \notag \\
  &\quad + K(1+ic_3) \left( \left< W \right> - W \right) \ . \label{eq:CGLE_LGC}
\end{align}
The linear average $\left< W \right>$ constitutes the linear global 
coupling, $K(1+ic_3)$ is the complex coupling strength with parameters $K$ and 
$c_3$.
Numerical results are depicted in Fig.~\ref{fig:LGC_chimera}. The
one-dimensional cut of $\left| W \right|$ reveals that the
spatio-temporal dynamics are qualitatively the same as in the type I
chimeras found in the MCGLE, see Fig.~\ref{fig:typeI}c. Moreover, the
arrangement of local oscillators in the complex plane in
Fig.~\ref{fig:LGC_chimera}c resembles the configuration in case of the
MCGLE. Differences are that the incoherent domains in
Fig.~\ref{fig:LGC_chimera}a are not of oval shape and that one
observes a rather irregular oscillation in the modulus in the
one-dimensional cut, instead of a more periodic one. These states are
related to chimera states found in an ensemble of Stuart-Landau
oscillators with linear global coupling \cite{Sethia_PRL_2014}. We
described in Ref.~\cite{Schmidt_PRL_2015} that in case of the chimera state 
under linear global coupling
the mean-field oscillates
approximately harmonically and that the amplitude clusters and
type I chimeras found in our model under nonlinear global coupling
constitute idealized dynamics of the corresponding
patterns found under linear global coupling.

\clearpage

\section{Type II dynamics}
\label{sec:typeII}

\begin{figure}[t]
  \centering
  \includegraphics[width=85mm]{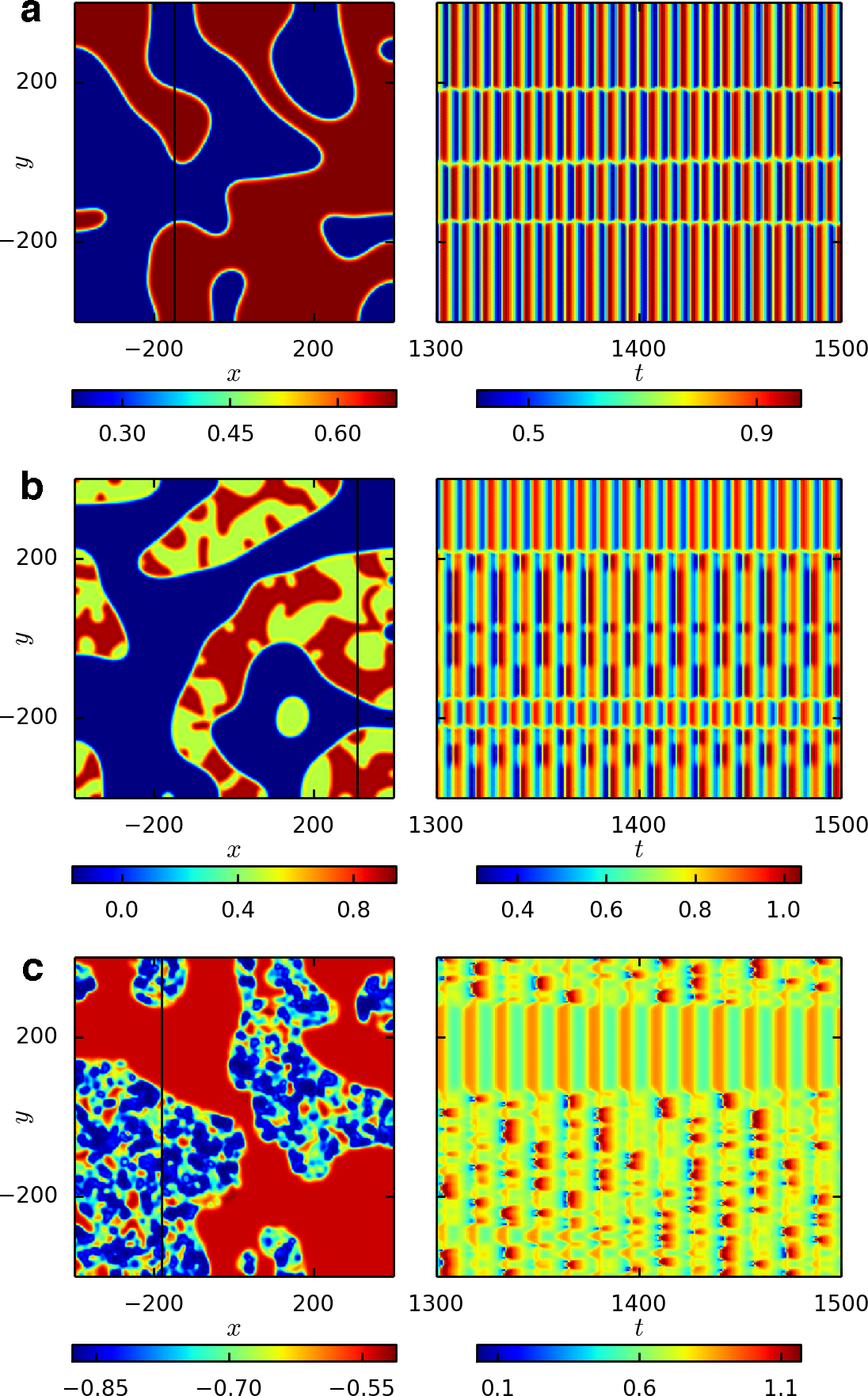}
  \caption{Type II patterns found in Eq.~\eqref{eq:MCGLE}:
    two-dimensional snapshots in left column (Re $W$), one-dimensional
  cuts in right column ($\left| W \right|$). (a) Modulated amplitude clusters ($c_1=0.2$, $c_2 = -0.7$, $\nu
  = 0.1$, $\eta = 0.66$). (b) Subclustering, where one phase is synchronized, while the other one
  exhibits two-phase clusters as a substructure ($c_1=0.2$, $c_2 = -0.67$, $\nu = 0.1$, $\eta =
  0.66$). (c) Type II chimera ($c_1=0.2$, $c_2 = -0.58$, $\nu = 0.1$, $\eta = 0.66$).}
\label{fig:typeII}
\end{figure}

Going back to the dynamics of Eq.~\eqref{eq:MCGLE}, for $c_2 = -0.7$,
$\nu = 0.1$ and $\eta = 0.66$, we find a second
type of cluster pattern, which is presented in
Fig.~\ref{fig:typeII}a. Note that in the snapshots we plot the real part of $W$,
as both the modulus and the phase exhibit significant variations in
case of the type II dynamics. The system again splits into two phases, but the
one-dimensional cut shows that the dynamics
are more complex than in case of amplitude clusters. The overall
uniform oscillation is modulated by two-phase
clusters. These modulational oscillations are visible in the
one-dimensional cut in Fig.~\ref{fig:typeII}a. Thus, we call this type of clustering modulated amplitude
clusters, as the clusters form in the modulational oscillations of the
amplitude. This pattern is one of the most prominent
patterns in the photoelectrodissolution of n-type silicon
\cite{Miethe_PRL_2009, GarciaMorales_PRE_2010, Schmidt_Chaos_2014,
  Schoenleber_NJP_2014} and the MCGLE thus reproduces this pattern
very well. 

The second type of clusters also undergoes a symmetry-breaking
transition \cite{Schmidt_Chaos_2014}, here when changing $c_2$ to $c_2 = -0.67$, resulting in
subclustering as shown in Fig.~\ref{fig:typeII}b. One phase exhibits two phase clusters as a
substructure, while the other one stays homogeneous. The
substructure-clusters oscillate at half the frequency of the
modulational oscillation. Therefore, we suspect the subclustering being
connected to a period-doubling bifurcation.

Changing the parameter further to $c_2 = -0.58$, the symmetry-breaking
becomes again more dramatic: the beforehand existing two-phase
subclusters turn into turbulence, thus realizing a second type of
chimera state, see Fig.~\ref{fig:typeII}c. Again we have to emphasize
the connection to the photoelectrodissolution of silicon, since this second
type of chimera states could be observed in experiments, too
\cite{Schmidt_Chaos_2014, Schoenleber_NJP_2014}. We compared the
simulations and the experiments in detail in
Refs.~\cite{Schmidt_Chaos_2014} and \cite{Schmidt_chapter_ECCII}. As
discussed in Ref.~\cite{Schmidt_Chaos_2014} the hierachy of patterns is
similar to the situation encountered in an experimental realization of
a system composed of two distinct groups with different intra- and
inter-group coupling \cite{Tinsley_Nature_2012}. In our case, the two
groups, i.e., the dynamically different phases, are created by the clustering mechanism and a difference
between intra- and inter-group coupling can be attributed to the phase
difference between the groups.

\begin{figure}[t]
  \centering
  \includegraphics[width=85mm]{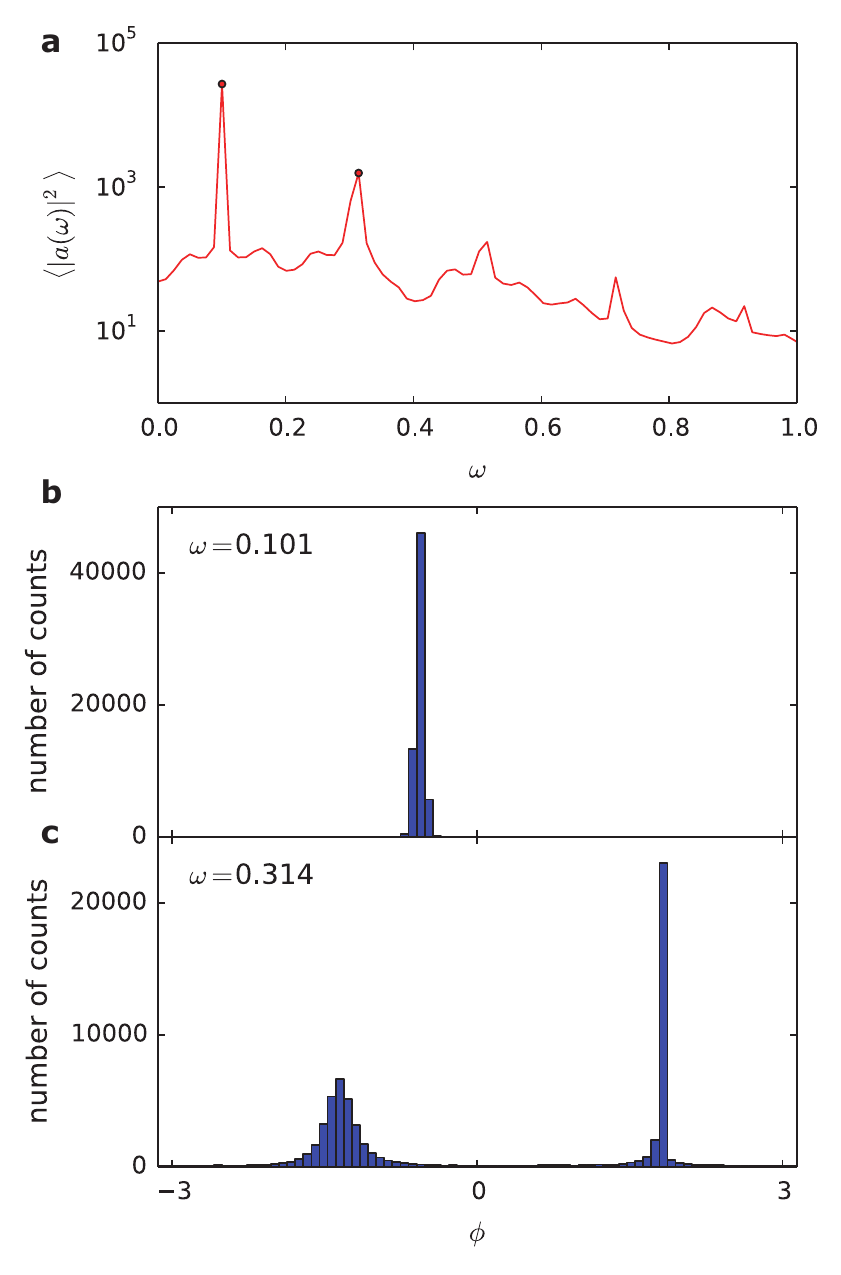}
  \caption{Spatially averaged power spectrum in (a) and phase histograms for
    the Fourier amplitudes $a(\mathbf r,\omega)$ at the highest peak in (b) and
    at the second highest peak in (c), which are indicated by circles
    in (a).}
\label{fig:Scum_histograms}
\end{figure}

In order to gain more insight into the spatio-temporal dynamics of the
type II chimera state, we analyze its frequency spectrum in a spatially
resolved manner. Therefore, we perform a Fourier transformation in
time of the real part of $W(\mathbf r,t)$ at every point $\mathbf
r=(x,y)$. We spatially average the resulting squared
amplitudes $\left| a(\mathbf r,\omega) \right|^2$ to obtain the
spatially averaged power spectrum $S(\omega) = \left< \left| a(\mathbf
    r,\omega) \right|^2 \right>$. The resulting spectrum for the
simulation presented in Fig.~\ref{fig:typeII}c is depicted in
Fig.~\ref{fig:Scum_histograms}a. Due to the turbulence in the
incoherent phase it consists of a large background, but exhibits also
two major peaks, marked with circles. The highest peak is at the
frequency $\nu$ of the mean-field oscillation, see
Eq.~\eqref{eq:MCGLE_avg}. The second highest peak stems from the
clustering frequency: as outlined above, in the modulated amplitude
clusters one observes two major frequencies, one of the homogeneous
oscillation and one as a result of the modulational oscillation. In the type II
chimera, the separation into two different phases occurs via this
clustering mechanism and therefore, some properties of it are still
present. This becomes more clear, when inspecting the Fourier
amplitudes $a(\mathbf r,\omega)$ at the two highest peaks, which
consist of moduli and phases. The distribution of phases of the
local oscillators at the two peaks is shown in histograms in
Figs.~\ref{fig:Scum_histograms}b and c. At frequency $\nu$,
Fig.~\ref{fig:Scum_histograms}b, the existence of only one, sharp peak
demonstrates that all local oscillators perform the homogeneous oscillation
in synchrony. The histogram for the second highest peak at $\omega
\approx 0.31$ (Fig.~\ref{fig:Scum_histograms}c) reveals that the clustering mechanism is still active:
we encounter two peaks for the two groups, which are phase shifted by
$\pi$, indicating that at this frequency the clustering occurs. The
right peak is sharp and describes the synchronized oscillators, while
the left peak has a Gaussian-like shape that stems from the
incoherent oscillators.

In essence, we encountered two types of cluster states giving rise to
two types of chimera states. A better understanding of these dynamics
can be obtained when omitting the diffusional coupling, thereby
reducing the complexity of the system. In the next section, we will do
so and demonstrate that diffusion is non-essential for the
dynamics presented so far.

\section{Dispensability of diffusion for type I and II dynamics}

\begin{figure}[h]
  \centering
  \includegraphics[width=85mm]{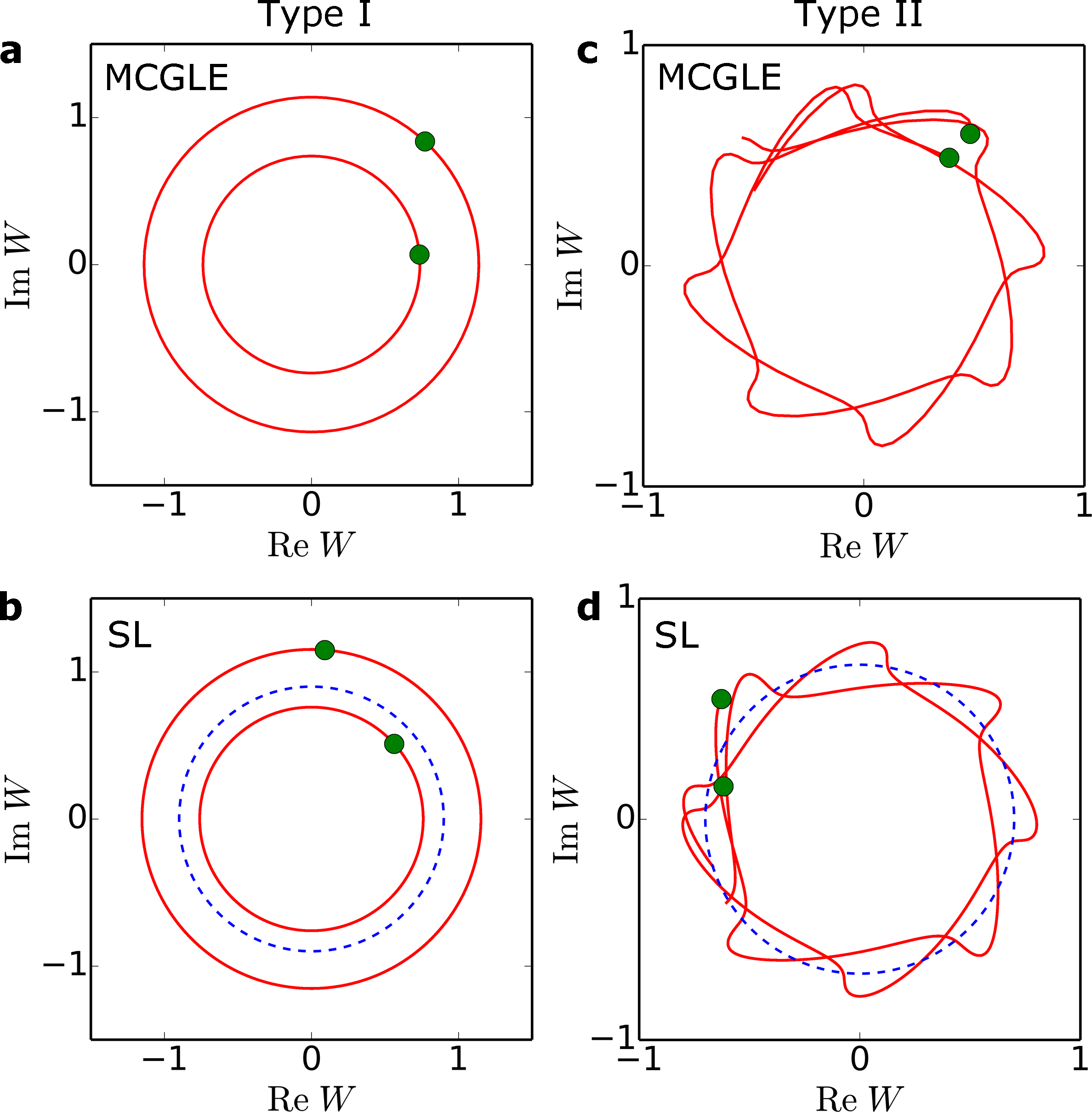}
  \caption{Cluster dynamics of individual oscillators in the MCGLE and
    in the Stuart-Landau ensemble (SL).
    (a) Amplitude cluster in the
    MCGLE ($c_1 = 0.2$, $c_2=-0.56$, $\nu=-1.5$, $\eta=0.9$). (b)
    Amplitude cluster in the Stuart-Landau ensemble ($c_2=-0.56$,
    $\nu=-1.5$, $\eta=0.9$).
    (c) Modulated amplitude cluster in
    the MCGLE ($c_1 = 0.2$, $c_2=-0.6$, $\nu=0.1$, $\eta=0.7$). (d)
    Modulated amplitude cluster in the Stuart-Landau ensemble
    ($c_2=-0.6$, $\nu=0.1$, $\eta=0.7$). In (b,d) the dashed line describes the 
    mean-field. Reprinted (d) with permission from [L. Schmidt \&
    K. Krischer, Phys. Rev. E \textbf{90}, 042911 (2014)]. Copyright
    (2014) by the American Physical Society (DOI: \href{http://dx.doi.org/10.1103/PhysRevE.90.042911}{10.1103/PhysRevE.90.042911}).}
  \label{fig:MCGLEvsSL_comp_cluster}
\end{figure}

The spatial coupling in the MCGLE, Eq.~\eqref{eq:MCGLE}, is given by the local
diffusional coupling and the nonlinear global coupling. Here, we
compare the cluster and chimera states of the MCGLE with dynamics
found in an ensemble of oscillators subject to nonlinear global coupling only \cite{Schmidt_PRL_2015}.
Therefore, we omit the diffusional coupling in the MCGLE and
consider an ensemble of Stuart-Landau oscillators with the nonlinear
global coupling:
\begin{align}
  \frac{\mathrm dW_k}{\mathrm dt} = &W_k - (1+i c_2)|W_k|^2 W_k \notag \\
  & - (1+i \nu)\left< W \right>_\Sigma + (1+i c_2)\left< |W|^2 W \right>_\Sigma \ .
  \label{eq:SL_ensemble}
\end{align}
Here, $k=1,2,\dots,N$ and $\left< \cdots \right>_\Sigma$ denotes the ensemble average, i.e.,
$\left< W \right>_\Sigma = \sum_{k=1}^N W_k/N$.
We compare the 
dynamics of oscillators in this Stuart-Landau ensemble with the 
dynamics of individual oscillators, or points in space, in the 
extended model, the MCGLE. Therefore, we choose a few points in 
space in each phase of the spatio-temporal patterns. For the cluster 
dynamics the resulting time series are depicted in 
Fig.~\ref{fig:MCGLEvsSL_comp_cluster}. The evolution in the complex 
plane in case of amplitude clusters in the MCGLE (a) is 
indistinguishable from the corresponding dynamics of oscillators in 
the Stuart-Landau ensemble (b). The same holds for modulated amplitude clusters 
in the MCGLE (c) and in the ensemble (d). A similar conclusion 
could be made for cluster patterns during the CO oxidation on Pt(110): 
Falcke \& Engel demonstrate that clusters arise with the global 
coupling alone, no local coupling is needed \cite{Falcke_PRE_1994}. In 
their experiment, the local coupling is due to surface diffusion of 
mobile adsorbates and the global coupling is due to pressure changes 
in the gas phase.

\begin{figure*}[h!]
  \centering
  \includegraphics[width=\textwidth]{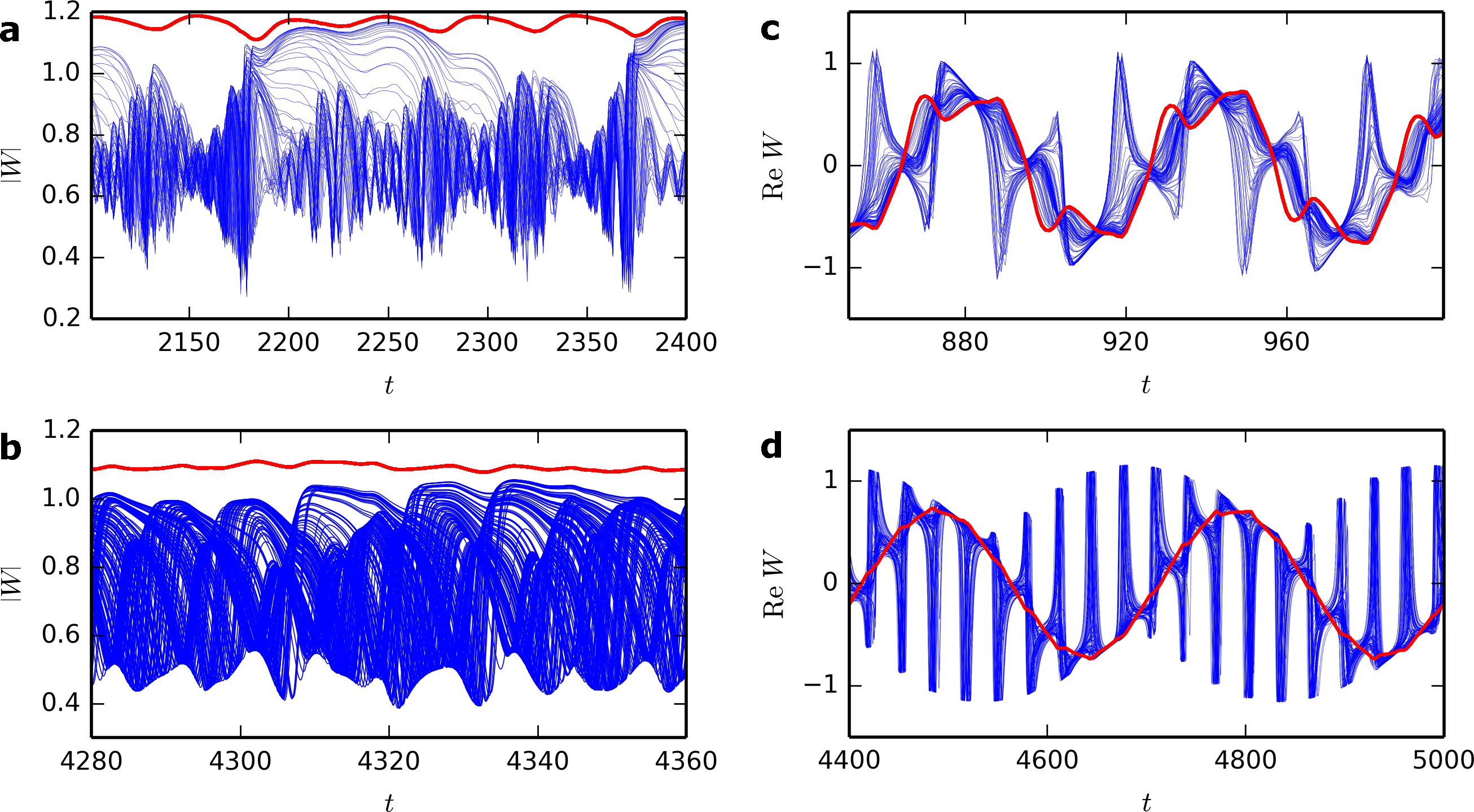}
  \caption{Dynamics in the two types of chimera states of individual
    oscillators in the MCGLE and in the Stuart-Landau ensemble. Red
    denotes the synchronized group and blue the desynchronized one. (a)
    Type I chimera in the MCGLE ($c_1 = 0.2$, $c_2=0.61$, $\nu=1.5$, $\eta=1.0$). (b) Type I chimera in the
    Stuart-Landau ensemble ($c_2=0.58$, $\nu=1.49$, $\eta=1.021$). (c)
    Type II chimera in the MCGLE ($c_1 = 0.2$, $c_2=-0.58$, $\nu=0.1$,
    $\eta=0.66$). (d) Type II chimera in the Stuart-Landau ensemble ($c_2=-0.6$, $\nu=0.021$, $\eta=0.7$).}
  \label{fig:MCGLEvsSL_comp_chimeras}
\end{figure*}

In Fig.~\ref{fig:MCGLEvsSL_comp_chimeras} the temporal dynamics in the two 
types of chimera states are considered. Individual oscillators in the 
MCGLE in case of type I chimeras (a) exhibit the same qualitative 
features as oscillators in the discrete ensemble (b): the synchronized 
group has a larger modulus, while the incoherent oscillators show 
irregular motion at lower values of the modulus. Note that the $c_2$, 
$\nu$ and $\eta$ values differ a bit between the extended and the 
discrete model. This is attributable to the change of stability by the 
diffusional coupling, i.e., in the extended system the chimera state 
spontaneously emerges at other parameter values as the diffusional 
coupling shifts the stability of the solutions. This might also be the 
reason why type I chimeras seem to be stable in the MCGLE, while 
we could only find unstable type I chimeras (undergoing heteroclinic transitions) in the Stuart-Landau 
ensemble \cite{Schmidt_PRL_2015}. However, on a qualitative basis the discrete model describes 
the same type I chimeras as the MCGLE.  

This holds also for type II chimeras, which are 
compared in Figs.~\ref{fig:MCGLEvsSL_comp_chimeras}c and d, for the 
MCGLE and the Stuart-Landau ensemble, respectively. In the MCGLE the 
oscillation of the synchronized group is not as harmonic as in the 
discrete model; the modulational oscillations are far more 
pronounced. The reason for this difference is that in the discrete 
model the desynchronized group is much smaller than the synchronized 
group, while in the extended system we observe approximately phase 
balance, i.e., both groups are of the same size. In both cases the 
mean-field has to oscillate harmonically. Hence, the spiking of the 
incoherent oscillators has to be compensated by the synchronized 
group, which is more pronounced if the groups are of the same size. 
Furthermore, due to the diffusional interaction of nearby oscillators, 
the spiking is smoother in the MCGLE. Again, the parameters are a 
bit different for the same reason as in case of type I chimeras.

Besides minor differences in the details of the dynamics, we can 
conclude that also the dynamics in the two types of chimera states in 
the MCGLE are sufficiently captured by the discrete Stuart-Landau 
ensemble. 

In the MCGLE, a bifurcation analysis of the cluster states is very
involved due to the high number of degrees of freedom. With the
knowledge that the diffusional coupling is dispensable, the problem highly
simplifies. In Ref.~\cite{Schmidt_PRE_2014} we considered two-cluster
solutions in the Stuart-Landau ensemble and performed a bifurcation
analysis. It turned out that starting with the synchronized solution the modulated amplitude clusters arise in
a secondary Hopf bifurcation, giving rise to a motion on a torus. The
two frequencies are then given by $\nu$ (frequency of the homogeneous
mode) and the frequency of the secondary Hopf bifurcation. Amplitude clusters are
then created in a saddle-node of infinite period bifurcation, thereby
destroying the torus. The two solutions of the amplitude clusters can
merge in a pitchfork bifurcation with the synchronized solution. For
more details see Ref.~\cite{Schmidt_PRE_2014}.

We can conclude that several complex spatio-temporal dynamics can be
described without diffusion. This enables one to analyze the dynamics
in more detail. However, this is not true for all solutions of Eq.~\eqref{eq:MCGLE}. In the next section we consider spatio-temporal dynamics that rely on a diffusional coupling.

\clearpage

\section{Localized turbulence}

\begin{figure*}[t]
  \centering
  \includegraphics[width=174mm]{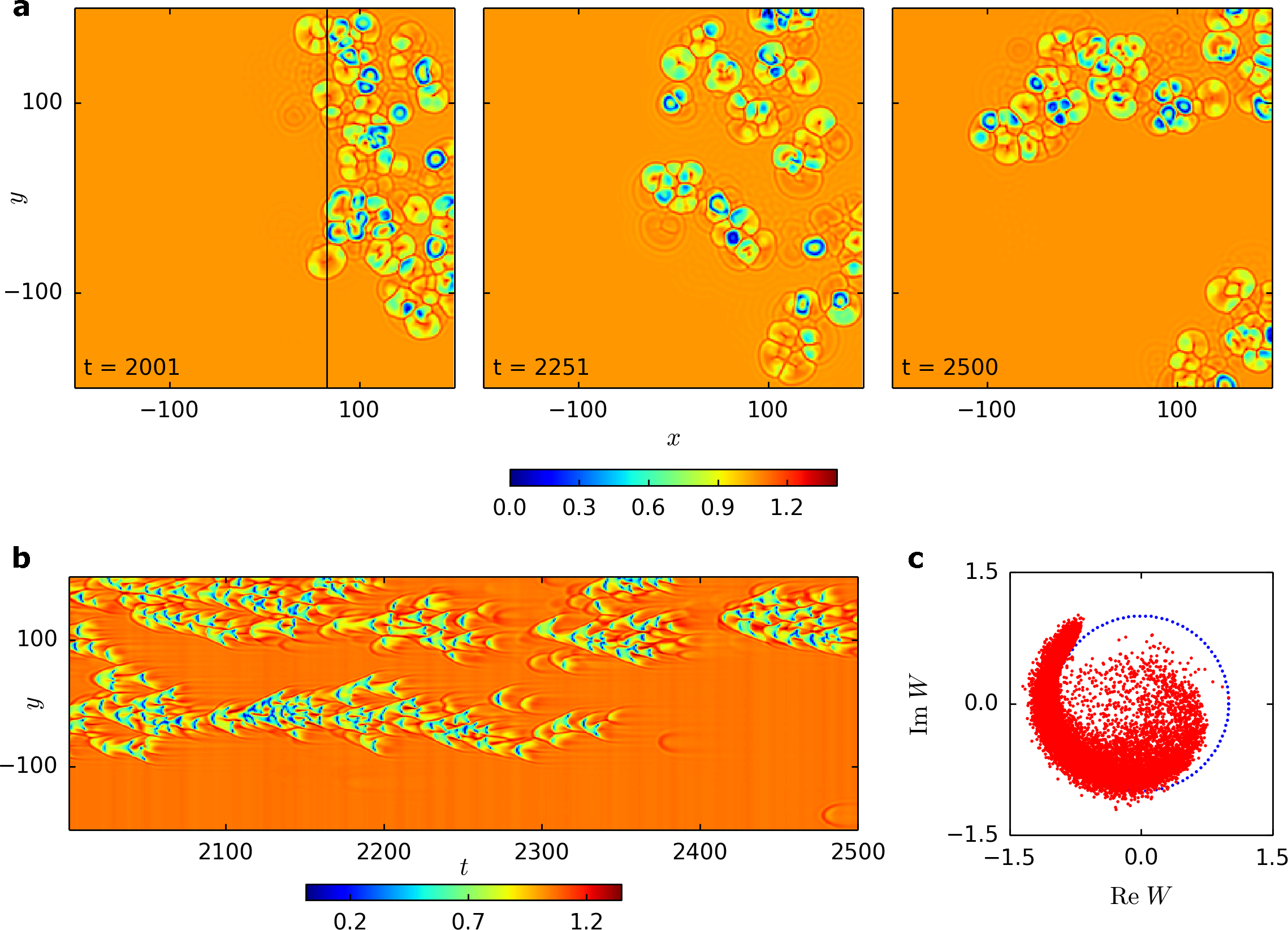}
  \caption{Localized turbulence in the MCGLE: two-dimensional snapshots of $\left|
      W \right|$ in (a) at times indicated in the figures, one-dimensional
    cut (also $\left| W \right|$) in (b) as indicated in the first snapshot and snapshot of the arrangement of local oscillators in
    the complex plane in (c) ($c_1=-1.6$, $c_2 = 1.5$, $\nu = 1.5$, $\eta = 0.9$).}
\label{fig:localized_turb}
\end{figure*}

For a totally different set of parameters, $c_1=-1.6$, $c_2 = 1.5$,
$\nu = 1.5$ and $\eta = 0.9$, we encounter so-called localized turbulence. We
depict our results in Fig.~\ref{fig:localized_turb}, where we present three snapshots at
times indicated in the figures in (a), a one-dimensional cut versus 
time in (b) and a snapshot of the arrangement of local oscillators in the complex
plane in (c). Here, we show again $\left| W \right|$. 
Small spots of
turbulence are randomly created in an otherwise homogeneous background, which then move through
the system and spread, but always keeping an overall small size. As
for their creation, they also vanish in an irregular manner. The
turbulence can be described as circular defect lines (or at least
lines of very small amplitude) expanding and breaking down. After their
breakdown, new defect circles are created. 
The mean extent of the turbulent region grows continuously when
changing parameters such that we go 
from a synchronized state to a fully turbulent state. Thus, localized
turbulence is a transitional state between synchronuous oscillations
and turbulence.
It has also been observed in the CGLE
with delayed global coupling \cite{Battogtokh_PhysicaD_1997}.

Localized turbulence is a phenomenon similar to chimera states as the system also shows
the coexistence of synchronized and turbulent regions. However, the
separation of these regions is fundamentally different. In the above
chimera states and also those discussed in 
literature, there is a clear separation of coherent and incoherent 
regions via a well-defined boundary, as clearly
visible in Fig.~\ref{fig:typeI}c, Fig.~\ref{fig:LGC_chimera} and
Fig.~\ref{fig:typeII}c. Also in nonlocally coupled systems, where
the boundaries are rather smooth, the separation is clearly
identifiable \cite{Kuramoto_NPCS_2002}. Furthermore, the boundaries move only on a slow
timescale. In contrast, the incoherent spots in localized turbulence 
are not separated from the homogeneous regions by phase boundaries. 
Hence, we face here the situation that although there is a coexistence of synchrony and incoherence, 
matching formally this most simple definition of a chimera state, 
the two phases, i.e., the 
synchronized and the incoherent phases, cannot be clearly identified. In our view, this
suggests
that localized turbulence should be discriminated
from a chimera state. However, this would necessitate a somewhat more restrictive
and perhaps more rigorous definition of a chimera state than we have it today.

\section{Conclusions}

With this study we widened the settings under which chimera states
exist: for each type of cluster state we observe, there exists a
corresponding chimera state. Since cluster formation has been studied
in various experimental setups, one can expect much more experimental
chimera states in the future.

We presented numerical results on the dynamics of an oscillatory
medium with a nonlinear global coupling that leads to preserved,
harmonic mean-field oscillations. It is a general model that captures
the dynamics observed during the photoelectrodissolution of n-type silicon:
the type II patterns
presented in Section~\ref{sec:typeII} reproduce experimentally found
spatio-temporal dynamics very well.

Two types of cluster states undergo a symmetry-breaking transition
towards two types of chimera states. Both chimera states inherit
properties from the cluster states, from which they originate. In case of
type I dynamics, the separation of the two groups is mainly by an
amplitude difference, for both clusters and chimeras. Type II
chimeras exhibit a phase shift of $\pi$ between the synchronized and
the desynchronized group at the clustering frequency, which is 
determined by the modulational oscillations. Thus, the two groups are 
created by the same clustering mechanism that gives also rise to the modulated 
amplitude clusters.
These considerations are in accordance with recent work on the discrete counterparts of the chimera
states in the Stuart-Landau ensemble in Ref.~\cite{Schmidt_PRL_2015},
which is reasonable, as we demonstrated that the diffusional coupling
is a non-essential ingredient. In fact, most of the dynamics can be
understood considering only the nonlinear global coupling, thereby
reducing the complexity of the system. As outlined in an earlier work,
a bifurcation analysis of the cluster states becomes possible in a
two-groups reduction of the Stuart-Landau ensemble
\cite{Schmidt_PRE_2014}. In addition, chimera states are found in an
ensemble of Stuart-Landau oscillators with linear global coupling
\cite{Sethia_PRL_2014} and there is a connection between these chimera
states and the type I chimeras in our system: Type I chimeras in the
Stuart-Landau ensemble with nonlinear global coupling constitute an
idealized case of the chimeras found under linear global coupling
\cite{Schmidt_PRL_2015}. As the discrete dynamics are the essential
dynamics, incorporating diffusion yields the same situation, as we
demonstrated in the present article. Thus, type I chimeras in the MCGLE and
chimera states found in a CGLE with linear global coupling are of the
same class. So far, we could not observe modulated amplitude clusters
under linear global coupling and consequently, no type II chimera
states are found. We suspect that the secondary Hopf bifurcation,
leading to modulated amplitude clusters, cannot occur with linear
global coupling.

Furthermore, we showed that the MCGLE also exhibits localized turbulence, a phenomenon for which 
diffusion seems to be necessary. In this state, turbulent spots move 
in an irregular manner through an otherwise homogeneous system. These 
spots are randomly created and destroyed. Because of the lack of a 
phase boundary between the turbulence and the homogeneous regions, we 
question that the term chimera state is appropriate for the
phenomenon of localized turbulence.

In general, a systematic classification of chimera states, as well as
a precise definition would be desirable. This requires an
understanding of how far it is reasonable to restrict the definition
such as: The system separates into two well-defined groups, one
synchronized, one incoherent, and each oscillator stays in one group
over times orders of magnitudes longer than typical oscillation
periods. For extended systems one might add that, after initial
transients, only oscillators in the boundary change the group. This
would provide a profound criterion, determining whether localized
turbulence is a chimera state or not.

% If in two-column mode, this environment will change to single-column format so that long equations can be displayed. 
%\begin{widetext}
%$$\mbox{put long equation here}$$
%\end{widetext}

% Figures should be put into the text as floats. 
% Use the graphics or graphicx packages (distributed with LaTeX2e).
% Put the label that you will use with \ref{} command in the braces of the \label{} command.
%
% \begin{figure}
% \end{figure}

% \begin{table}
% \end{tabular}
% \end{table}

\begin{acknowledgments}
We thank S.~W. Haugland for fruitful discussions.
Financial support from the \textit{Deutsche Forschungsgemeinschaft}
(grant no. KR1189/12-1),
the \textit{Institute for Advanced Study - Technische Universit\"{a}t
  M\"{u}nchen} funded by the German Excellence Initiative and the cluster of excellence \textit{Nanosystems
  Initiative Munich (NIM)} is gratefully acknowledged.
\end{acknowledgments}

\appendix
\section{}
\subsection{Simulations of the extended systems}
Simulations of Eqs.~\eqref{eq:MCGLE} and \eqref{eq:CGLE_LGC} in the main text are carried out using a
pseudospectral method and an exponential time stepping algorithm
\cite{CoxMatthews_JCompPhys_2002}. We use 512x512 Fourier modes, a
computational timestep of $\Delta t = 0.05$ and
system sizes as indicated in the figures. Note that the equation is
dimensionless. The system is initialized with a two-dimensional
circular perturbation and additional noise. Independent runs with
different seeds produce the same qualitative behavior. This even
holds, if we consider only random initial conditions without the
circular perturbation.

\subsection{Simulations of the discrete ensemble}
We solved Eqs.~\eqref{eq:SL_ensemble} using an implicit Adams method
with a timestep of $dt = 0.01$. Initial conditions are random numbers
on the real axis fulfilling the conservation law for the mean-field.

\bibliography{lit}

\end{document}